\documentclass[
aps,%
12pt,%
final,%
notitlepage,%
oneside,%
onecolumn,%
nobibnotes,%
nofootinbib,%
superscriptaddress,%
noshowpacs]%
{revtex4}%
\usepackage{graphics}
\usepackage[T2A]{fontenc}
\usepackage[english]{babel}

\begin{document}

\title{Quark-hadron duality and production  of charmonia and 
doubly charmed baryons in  $e^+e^-$ annihilations}
\author{\firstname{A.~V.}~\surname{Berezhnoy}}
\email{aber@ttk.ru}
\affiliation{SINP MSU, Moscow, Russia}%
\author{\firstname{A.~K.}~\surname{Likhoded}}
\email{likhoded@mx.ihep.su}
\affiliation{IHEP, Protvino, Russia}

\begin{abstract}
In the framework of quark-hadron duality, the cross section values of 
charmonium and doubly charmed baryon production in  $e^+e^-$-annihilations
have been estimated at interaction energy $\sqrt{s}=10.6$~GeV. 
It has been shown that the approach under discussion
allows to resolve some contradictions between the experimental data 
and pQCD predictions.
\end{abstract}

\maketitle

\section{Introduction}

\begin{figure} 
{\centering \resizebox*{\textwidth}{!}{\includegraphics{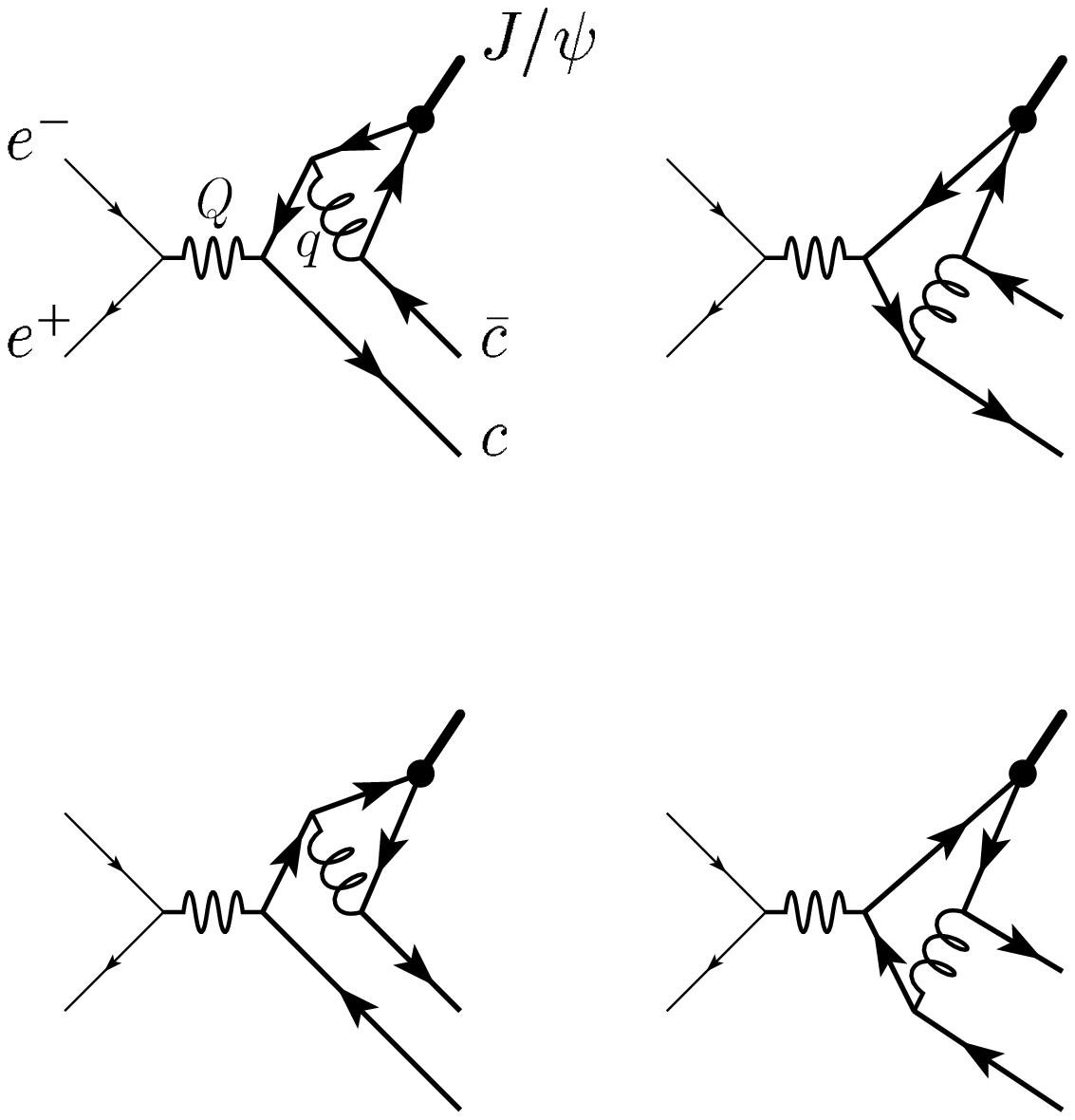}}
\vspace*{-4cm} \par}

\vspace*{-6cm}
\caption{The leading order diagrams for the process $e^+e^-\to J/\psi +c \bar
c$. $Q$ is  momentum of the photon and 
$q$ is momentum of the intermediate quark or gluon.  The average  of $q^2$
in the total phase space is about several $m_c^2$.
If both $c\bar c$-pairs are fused into charmonia 
($e^-e^+\to J/\psi + \mbox{\rm charmonium}$), then $q^2\sim Q^2/4$ within the
$\delta$-approximation. 
\hfill}
\label{jpsicc_ccdual.ps}
\end{figure}

Until recent times, the calculation of $J/\psi$ production cross section   
was done within the so-called $\delta$-approximation, which is based on the
factorization theorem of QCD.  In the approach under discussion,  relative 
momenta of valence quarks of $J/\psi$-meson are not taken into account  in 
the hard part of amplitude. One supposes that both valence quarks are
on-shell and each of them carries a half of the total $J/\psi$-meson momentum.  
Nevertheless, the recent experimental results of the BELLE and BABAR 
Collaborations on 
$J/\psi$-production in $e^+e^-$ annihilations demand an essential revision
of the calculation techniques based on QCD factorization theorem.  
Indeed, the cross section value of charmonium pair production  
in $e^+e^-$ annihilations estimated in the framework of
the $\delta$-approximation~\cite{2charmonium} underestimates the experimental 
data by an order of magnitude~\cite{Belle, BABAR}.
The detailed analysis has shown that such underestimation is caused by
large fixed virtualites of the intermediate quark and gluon
$q^2\sim \frac{Q^2}{4}$ ($Q^2$ is a virtuality of the initial photon, 
see Fig.~\ref{jpsicc_ccdual.ps}),
which occur within the $\delta$-approximation technique.  

Taking into account 
the relative motion of valence quarks allows to remove contradictions
between theory and experiment, at least in principle. 
Calculations of the pair quarkonium production in $e^+e^-$ annihilations, 
 obtained  within a light cone formalism~\cite{lightcone}
are in a qualitative agreement with the experimental data  of the 
BELLE and BABAR Collaborations. In these works, the $\delta$-shaped wave 
function of quarkonium
was replaced by a  function  $f(x)$, which is  "spread" on $x$, 
where $x$ is a momentum fraction of the quarkonium carried 
by a valence quark in the infinite momentum frame. 
Thus, the longitudinal "internal motion" of the valence quarks
is taken  into account 
\footnote{It is worth to note that the problem of internal
motion  of valence quarks inside $J/\psi$-meson was discussed 
a rather long time ago.
 See, for example, works \cite{KLP_frag}, where it was  suggested to
use  the structure function of $J/\psi$ in form $F_c= x^{2.2}(1-x)^{2.45}$.}.  
This spreading leads to a decrease of effective virtuality 
of the intermediate quark and gluon,  and, therefore,  to  an 
increase of the predicted cross section value.

In spite of certain success in the description of 
charmonium pair production,   uncertainties of the concrete model
realization remain. 
These are uncertainties of wave function value, 
final $c$-quark virtualities, and standart uncertainties of
running constant value $\alpha_s$,  $c$-quark mass value $m_c$, etc.

The  cross section of  inclusive charmonium production in  
$e^+e^-$ annihilations ($e^+e^- \to \mbox{\textrm{charmonium}} +c+\bar c$)  
depends on model uncertainties to a lesser extend~\cite{Leibovich,Jpsi_cc}, 
because  the largest cross section fraction for such process is 
determined by the smallest virtualities of the intermediate quark and gluon 
(Fig.~\ref{jpsicc_ccdual.ps}).
Taking into account the relative motion of quarks changes the 
the result less than 10\%~\cite{Nobary}, and  a general dependence
on choice of $m_c$ and $\alpha_s$ values remains. 

In this work we perform a new analysis  of 
four charmed quark production in $e^+e^-$ annihilations within the
quark-hadron duality.

\section{Exclusive and inclusive charmonium production}
It is reasonable to describe the cross section of four 
 $c$-quark production in the process 
\begin{equation}
e^+e^-\to c\bar cc\bar c, 
\label{4c}
\end{equation}
in the framework of perturbative QCD.  A value of
$m_c=1.25$~GeV can be  chosen for the $c$-quark mass. This  value is the same
as that adopted for QCD sum rules.  
 The strong coupling constant at the scale of $m_c$ 
is put to  $\alpha_s=0.24$.  It is worth to mention that the
choice of relatively small value of $c$-quark mass makes a use of
quark-hadron duality  more convincing, because the masses
of observed doubly heavy hadrons lie approximately in the center of 
duality interval at this $m_c$ value. 

The cross section value of process  (\ref{4c}) 
can be considered as a sum  of  cross section
values of  all process with four  $c$-quarks. 
Leading order calculations with the chosen values of  $\alpha_s$ and
$m_c$  at  $\sqrt{s}=10.6$~GeV give the following 
cross section value  of the process 
\begin{equation}
\sigma(e^+e^-\to c\bar c c \bar c)=372\,\mbox{fb}.
\label{sig_4c}
\end{equation} 
The cross section of pure electro-dynamical production of
four $c$-quarks  is only  $1\div 2$\% of the above value. 

There are uncertainties in prediction for  the process (\ref{4c})
due to  the choice of  $m_c$ and  $\alpha_s$ values.
The $m_c$ and  $\alpha_s$  used here lead to the most optimistic 
prediction for the cross section value of four $c$-quark production.
This value can be considered as an upper limit  of  four $c$-quark yield. 
The performed estimation (\ref{sig_4c}) 
is in accordance with the  cross section value at  $Z^0$-boson peak~\cite{Z0}. 

As it was noted earlier~\cite{Xi_cc}, for  the  processes 
 \begin{equation} 
e^+e^-\to J/\psi +c+\bar c, 
\label{Jpsi_delta}
\end{equation}
\begin{equation} 
e^+e^-\to \eta_c +c+\bar c, 
\label{eta_delta}
\end{equation}
the total cross section values estimated within the
$\delta$-approximation~\cite{Leibovich,Jpsi_cc}  agree with
the cross section value  of production of 
color singlet  $c\bar c$-pair calculated for a some duality interval. 

Using values of wave function at origin $|R_s (0)|^2=0.67\,\mbox{GeV}^3$ and
the charmonium mass value 3~GeV, the following results have been obtained 
for  $J/\psi$  and $\eta_c$ production at $\sqrt{s}=10.6$~GeV:
\begin{equation}
\sigma (e^+e^-\to J/\psi +c+\bar c)=104\,\mbox{fb},
\label{Jpsi_delta_sig}
\end{equation}
\begin{equation}
\sigma (e^+e^-\to \eta_c +c+\bar c)=40\,\mbox{fb}.
\label{eta_delta_sig}
\end{equation}
Therefore,
\begin{equation}
\sigma (J/\psi)/\sigma (\eta_c ) =2.6.
\end{equation}

A contribution of higher  $S$- and $P$-wave charmonium excitations  into the 
inclusive production cross section is about  50\%~\cite{Jpsi_cc},  
consequently the total cross section value 
can be estimated\footnote{Here and below the process of one-photon 
 annihilation is discussed.
Two-photon annihilation $e^+e^-\to \gamma^* \gamma^* \to J/\psi +c+\bar c$
contributes less than 10\%  
into the total cross section value~\cite{twophotons}.} as 
 216~fb. 
This value can be  compared to the cross section value 
obtained in the appropriate duality interval of mass spectrum of 
color singlet $c\bar c$-pairs produced in the process  (\ref{4c}):
\begin{equation}
2m_c<m_{c\bar c}<2m_{D}+\Delta,
\label{interval}
\end{equation}
where  $m_c=1.25$~GeV, $\Delta=0.5$~GeV, and  $m_{D}$ is the experimental
$D$-meson mass.
Thus, one can obtain:

\begin{equation}
\left.\int_{2m_c}^{2m_{D}+\Delta} 
\frac{d\sigma\left(e^+e^-\to (c\bar c)_{\rm{singlet}}+c+\bar c \right)}
{dm_{c\bar c}}
dm_{c\bar c}\right|_{\Delta=0.5 \,\mbox{\rm GeV}}=280\, \mbox{\rm fb}.
\end{equation}
This value  should be compared with 216~fb.  
Therefore,  for the case of inclusive charmonium production 
(\ref{Jpsi_delta}) and (\ref{eta_delta}) for some
$\Delta$  a fair agreement between the results of $\delta$-approximation and 
the quark-hadron duality hypothesis can be achieved. 
For the color singlet  pairs with a definite spin value 
($S=0$ or $S=1$) one can obtain:
\begin{equation}
\left.\int_{2m_c}^{2m_{D}+\Delta} 
\frac{d\sigma(e^+e^-\to (c\bar c)^S_{\rm{singlet}}+c+\bar c)}{dm_{c\bar c}}
dm_{c\bar c}\right|_{\Delta=0.5 \,\mbox{\rm GeV}}^{ S=1}=204\, \mbox{\rm fb,}
\end{equation}
\begin{equation}
\left.\int_{2m_c}^{2m_{D}+\Delta} 
\frac{d\sigma(e^+e^-\to (c\bar c)^S_{\rm{singlet}}+c+\bar c)}{dm_{c\bar c}}
dm_{c\bar c}\right|_{\Delta=0.5 \,\mbox{\rm GeV}}^{ S=0}=76\, \mbox{\rm fb.}
\end{equation}
If one supposes that the spin-triplet  $c\bar c$-pair ($S=1$)  most
probably transforms into a $J/\psi$ meson  and 
the spin singlet  $c\bar c$-pair ($S=0$)  most
probably transforms into $\eta_c$, then  
\begin{equation}
\sigma (J/\psi)/\sigma (\eta_c ) =2.7.
\end{equation}
This ratio is in a good agreement with one calculated in the frame 
work of the $\delta$-approximation. 

Thus for the case of inclusive charmonium production, 
the both  approaches under discussion agree.
Contrary, for a case of exclusive charmonium pair production the cross section 
value calculated within the quark-hadron duality hypothesis is larger than 
the $\delta$-approximation prediction by one order of magnitude.   
Such a difference occurs because the quark-hadron duality allows 
to take into account the internal motion of quarks inside charmonium. 
Indeed, a quark and an antiquark produced in duality interval (\ref{interval}) 
and fused into charmonium according to the quark-hadron duality hypothesis
can have different velocities. This
decreases the effective virtuality of the intermediate gluon and quark, 
and increases consequently  the predicted cross section value. 

The exclusive charmonium pair production cross section 
can be estimated by  picking out the 
both $c\bar c$-pairs produced  in  the process (\ref{4c}) in 
appropriate duality intervals and in color-singlet state.   
In our calculations, we will also distinguish spin states of the pairs 
($S_1$ and $S_2$, correspondingly). 
The calculation results are as follows:
\begin{equation}
\left. \int\!\!\!\int_{2m_c}^{2m_{D}+\Delta} 
\frac
{d^2\sigma\left(e^+e^-\to (c\bar c)^{S_1}_{\rm{singlet}}+(c\bar
c)^{S_2}_{\rm{singlet}}\right)}{dm_{{c\bar c}_1}dm_{{c\bar c}_2}}
dm_{{c\bar c}_1}dm_{{c\bar c}_2}
\right|_{\Delta=0.5 \,\mbox{\rm GeV}}^{ S_1=0, \, S_2=0}
=1.2\, \mbox{\rm fb,}
\end{equation}
\begin{equation}
\left. \int\!\!\! \int_{2m_c}^{2m_{D}+\Delta} 
\frac
{d^2\sigma\left(e^+e^-\to (c\bar c)^{S_1}_{\rm{singlet}}+(c\bar
c)^{S_2}_{\rm{singlet}}\right)}{dm_{{c\bar c}_1}dm_{{c\bar c}_2}}
dm_{{c\bar c}_1}dm_{{c\bar c}_2}
\right|_{\Delta=0.5 \,\mbox{\rm GeV}}^{ S_1=0, \, S_2=1}
=19.4\, \mbox{\rm fb,}
\label{pv}
\end{equation}
\begin{equation}
\left. \int\!\!\! \int_{2m_c}^{2m_{D}+\Delta} 
\frac
{d^2\sigma\left(e^+e^-\to (c\bar c)^{S_1}_{\rm{singlet}}+(c\bar
c)^{S_2}_{\rm{singlet}}\right)}{dm_{{c\bar c}_1}dm_{{c\bar c}_2}}
dm_{{c\bar c}_1}dm_{{c\bar c}_2}
\right|_{\Delta=0.5 \,\mbox{\rm GeV}}^{ S_1=1, \, S_2=1}
=19.8\, \mbox{\rm fb.}
\label{vv}
\end{equation}

One can suppose that if $S_1=0$ and  $S_2=1$ (or, $S_1=1$ and $S_2=0$), then 
  $J/\psi$ and $\eta_c$ are produced in  most cases; if
 $S_1=1$ and $S_2=1$, then
$J/\psi$ and $P$-wave state of quarkonium are most probably produced. 
Then the cross section of process
\begin{equation} 
e^+e^-\to J/\psi +\mbox{\rm charmonium} 
\label{Jpsi_charmonium}
\end{equation}
at $\sqrt{s}=10.6$~GeV can be estimated to be about 40~fb.
This value can be compared with the experimental data on the charmonium
pair production:
\begin{equation}
\sigma (e^+e^-\to J/\psi +\mbox{\rm charmonium})=55\pm 10\,\mbox{fb}\;\;\;\mbox{(BELLE Collaboration)},
\label{BELLE}
\end{equation}
\begin{equation}
\sigma (e^+e^-\to J/\psi +\mbox{\rm charmonium})=44.3\pm 9\,\mbox{fb}\;\;\;\mbox{(BaBar Collaboration)}.
\label{BaBar}
\end{equation}

One can conclude that the 
quark-hadron duality prediction  does not contradict the experimental data
within experimental errors and uncertainties in $m_c$ and  $\alpha_s$ values.
There is certaintly no reason to talk about the  discrepancy of order of
magnitude, as it is a case for the predictions of the
$\delta$-approximation~\cite{2charmonium}.
It is worth to note, that the calculation results  
(\ref{pv}) and (\ref{vv}) indicate a large yield of 
 $P$-wave states, which is comparable with the  yield 
 of  $\eta_c$ and  $\eta'_c$.  It is also in agreement with
 the experimental data of the BELLE and BaBar Collaborations.

In the work~\cite{color_evap} the quark-hadron duality hypothesis was
combined with the Colour Evaporation Model predincting
 a yield of charmonium state with 
the total angular momentum $J$ is proportional to  $2J+1$. 
Obviously, this model cannot be
applied to the process $e^+e^-\to J/\psi +X$ 
because the  quantum numbers of $J/\psi$ and  $X$ channel are strongly
correlated.

The above estimations  allow to
predict the cross section  of process $e^+e^-\to J/\psi+D\bar D$. 
Subtracting the cross section value
of exclusive pair production process from the cross section 
value of inclusive charmonium production process (\ref{Jpsi_delta}) one can
obtain:
\begin{equation}
\sigma
(e^+e^-\to J/\psi+D+\bar D)\simeq 240\,\mbox{fb}. 
\end{equation} 

That value can be compared with that obtained by  the
BELLE Collaboration~\cite{Belle-2}:
\begin{equation}
\sigma (e^+e^-\to J/\psi+c+\bar c)
=870\pm 26\,\mbox{fb}.
\end{equation}
Thus, our predictions underestimates the experimental results by
several times.
Moreover, within the model under discussion the cross section value 
for the process $e^+e^-\to J/\psi+D\bar D$ cannot be larger than
the cross section value of four charmed quark production, i.e. it can be not
large then 330~fb,
and the essential fraction of this  value 
should include  doubly charmed baryon contribution. 
It is discussed in the next chapter.

\section{Doubly charmed baryon production}
Let us estimate how often a $cc$-pair  is produced in the color 
antitriplet state with a small invariant mass in a system of four 
$c$-quarks. Such an object can be 
considered as a candidate to create a doubly charmed baryon 
(see also~\cite{Xi_cc}).
In the duality interval $2m_c< m_{cc}< 2m_D+\Delta$ for 
$\alpha_s=0.24$, $m_c=1.25$~GeV,  and $\Delta=0.5$~GeV,
one obtains the following cross section value:
\begin{equation}
\sigma \left(e^+e^-\to (cc)_{\bar 3}/(\bar c \bar c)_3+X 
\right)=170\,\mbox{fb}. 
\end{equation} 

For the production of both  $cc$- and $\bar c\bar c$-pairs 
in the above duality interval, the cross section value is
\begin{equation}
 \sigma \left(e^+e^- \to (cc)_{\bar 3}+(\bar c\bar c)_3\right)
 =32 \,\mbox{fb}. 
\end{equation} 

We assume that in the kinematical region  under discussion
a doubly charmed baryon and an antibaryon are
created with high  probability.
The rest of the cross section value 
$\sigma (e^+e^-\to (cc)_{\bar 3}/(\bar c \bar c)_3+X)$ (about 140~fb)
should be considered as an upper limit for 
the cross section of the processes 
like $e^+e^-\to\Xi_{cc}+\bar \Lambda _c+\bar D$.

Let us discuss here a relation between 
the inclusive charmonium production and the production of doubly 
charmed baryons. The  $c$-quark and $\bar c$-quark produced in
 the intermediate gluon splitting  have rather often almost the same
velocities and this  provides 
the minimum virtuality of intermediate gluon ($\sim 4 m_c^2$).
Let us suppose that the color singlet $c\bar c$-pair is produced
with a small invariant mass. Either $c$-quark or  
$\bar c$-quark in this pair is produced by the splitting gluon.
For the sake of simplicity, let us suppose that it is a $\bar c$-quark. 
Then the remaining $c$-quark 
(which does belong to the pair under consideration)
 is also produced by the splitting gluon. It means that
there is a large probability for this $c$-quark to have a velocity 
 close to the velocity of the color singlet $c\bar c$-pair. This is why
the remaining $c$-quark can create with $c$-quark from $c\bar c$-pair an 
antitriplet state with a small invariant mass.
Therefore such a $cc\bar c$-system can create  
a charmonium, as well as a doubly charmed baryon.   
Within the $\delta$-approximation, the relative inclusive yield of
charmonia and doubly charmed baryons  are determined by  the ratio of 
squared wave functions at origin for charmonium and doubly charmed diquark,
correspondingly.  Unfortunately, the quark-hadron duality technique
does not allow to definitely separate  these yields  for the case of 
inclusive production.

Contrary, the exclusive process of charmonuim pair production and the
process of pair production of doubly charmed baryons 
are well separated kinematically, and the problem under discussion does not
arise in this case.

\section{Process   \NoCaseChange{\(e^+e^-\to J/\psi gg\)} } 

\begin{figure} 
{\centering \resizebox*{\textwidth}{!}{\includegraphics{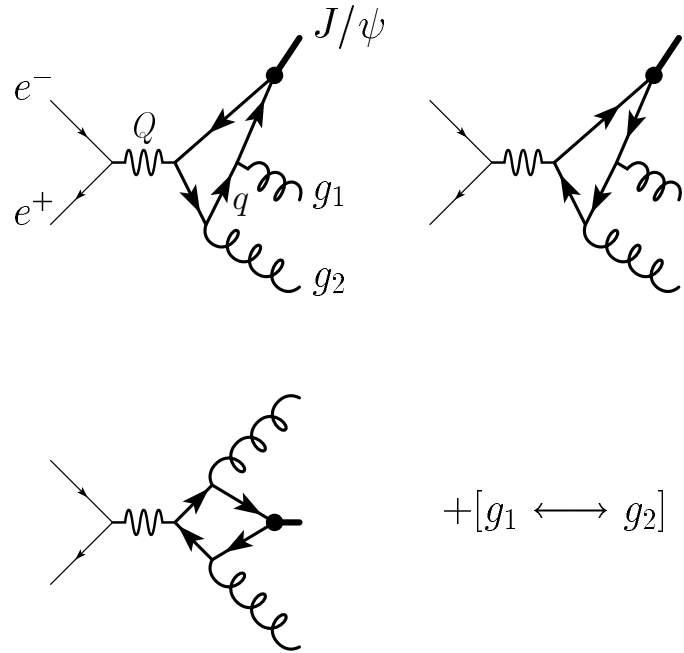}}
\vspace*{-4cm} \par}

\vspace*{-4cm}
\hspace*{2cm}
\caption{The leading order diagrams for the process $e^-e^+\to J/\psi gg$.} 
\label{jpsigg_ccdual.ps}
\end{figure}

When the  charmonia production in $e^+e^-$-annihilation
is discussed,  the process  
\begin{equation}
e^+e^-\to J/\psi gg 
\label{Jpsi_gg}
\end{equation}
cannot be ignored (see Fig.~\ref{jpsigg_ccdual.ps}). The detailed study 
of the process  is worthy of a separate
article, because the situation here is quite similar  to one with the exclusive
pair production of charmonia (\ref{Jpsi_charmonium}).
The treating of the process (\ref{Jpsi_charmonium}) within the
$\delta$-approximation 
leads to large fixed virtualities of intermediate quarks, as 
in the case of process (\ref{Jpsi_charmonium}).
This means, that the $\delta$-approximation leads to the underestimated 
value of the cross section of the process (\ref{Jpsi_gg}), as well as  
 (\ref{Jpsi_charmonium}). The value of effective virtuality 
predicted within the quark-hadron duality is less then  the value 
predicted within the $\delta$-approximation, and it leads to the cross section 
value increase.  In this work we do not intend to study the process
 $e^+e^-\to J/\psi gg$ in details and only give the 
cross section value  predicted in the framework of hypothesis of
the quark-hadron duality.   For $\Delta=0.5$~GeV and for the invariant 
mass of two final  gluons  larger than two masses of $\pi$-meson
 ($m_{gg}>2M_{\pi}$), one can obtain:
\begin{equation}
\left.\int_{2m_c}^{2m_{D}+\Delta} 
\frac{d\sigma(e^+e^-\to (c\bar c)^S_{\rm{singlet}}+gg)}{dm_{c\bar c}}
dm_{c\bar c}\right|_{\Delta=0.5 \,\mbox{\rm GeV}, \, m_{gg}>2M_{\pi}}^{S=1}
=1.96\, \mbox{\rm pb,}
\label{sig_Jpsi_gg}
\end{equation}
\begin{equation}
\left.\int_{2m_c}^{2m_{D}+\Delta} 
\frac{d\sigma(e^+e^-\to (c\bar c)^S_{\rm{singlet}}+gg)}{dm_{c\bar c}}
dm_{c\bar c}\right|_{\Delta=0.5 \,\mbox{\rm GeV}, \, m_{gg}>2M_{\pi}}^{S=0}
=1.00\, \mbox{\rm pb.}
\end{equation}

If one suppose that the pair with a spin value of $S=1$ 
 most probably creates  $J/\psi$-state, 
then the results (\ref{sig_Jpsi_gg}) should be considered as the cross 
section value estimation for the process $e^+e^-\to J/\psi gg $.
Of cause, this estimation is quite rough.  In addition, 
our calculations show that  the prediction value depends quite strongly on 
 $\Delta$. For example, at  $\Delta=0.1$~GeV the cross 
section value of the $e^+e^-\to J/\psi gg $ is about
0.6~pb, which is in agreement with the experimental data.
To achieve a better reliability  of the predictions,  one  can introduce 
an effective mass of gluon into calculations, 
as it was done in the works~\cite{Jpsi_gg_mass,Jpsi_gg_BL}. 
Therefore, the quark-hadron duality hypothesis does not contradict   the
experimental data. Contrary, 
the predictions of the
$\delta$-approximation~\cite{Jpsi_gg_BL,Jpsi_gg_delta}\footnote{
In the work~\cite{Jpsi_gg_BL} the factor of 1/2 has been missed in the cross
section calculation for the process $e^+e^-\to J/\psi gg $. 
 The revised prediction value is about $0.35$~pb instead of $0.7$~pb.}
 are too small to describe the data.

\section{Conclusions}

Until recently, it was believed that the experimental data on
$J/\psi$ production in  $e^+e^-$-annihilation cannot be 
described within the perturbative  QCD. This conclusion was based on
calculation which does not take into account the motion of the valence quark
inside the   $J/\psi$-meson.  Taking into account the
"internal motion" by means of light cone wave function, as it is 
done in works~\cite{lightcone}, or by means of 
quark-hadron duality, as it is done in the present work, allows 
 to resolve some contradictions between  the  pQCD and the experimental data.
One can certainly claim  that  the experimental data on the charmonium
 pair production in $e^+e^-$-annihilation, as well as the  experimental data on
charmonium production in the process 
$e^+e^-\to J/\psi gg $, do not contradict pQCD predictions.

Nevertheless,  the experimental value of the inclusive cross section of
 $J/\psi$ production in the process $e^+e^-\to J/\psi +D +\bar D$ is still
underestimates by theoretical predictions. Even our very optimistic estimation
of the cross section value (330~fb at 10.6~GeV) is approximately 
three times  less than the experimental result. Moreover, 
we think that the value of about 200~fb seems
 to be the more reasonable estimation
of the process cross section.

In conclusion it is worth to mention one more time, that the calculations
done in this work within the quark-hadron duality hypothesis
suggest,  that the cross section  of the pair production of
doubly charmed baryons in $e^+e^-$-annihilation  could be of the same 
order as the cross section  of charmonium pair production.

We thank L.~K.~Gladilin and S.~P.~Baranov for the frutfull discussion.
This work was partially supported by Russian Fund of Basic
Research (project  No~04-02-17530),  
and the CRDF (project No.~M0-011-0)  and  was performed within the
program for Support of Leading Scientific Schools (project  No.~1303.2003.2).
 The work of A.V.~Berezhnoy  was also supported in part by
Russian Fund of Basic Research (project No.~05-07-90292),
by President of Russian Federation (project for Young PhD Support
 No.~2773.2005.2), 
 and by Dynasty Foundation (project for Young Scientist Support).

\newpage

\end{document}